# Clustering Based Topology Control Protocol for Data Delivery in Wireless Sensor Networks

**Deepali Virmani**
**Department of CSE**
**BPIT, GGSIPU, Delhi, India**
deepalivirmani@gmail.com

**Satbir Jain**
**Department of CSE**
**NSIT, Dwarka, Delhi, India**
jain_satbir@yahoo.com

**Abstract.** The issue of optimizing the limited and often non-renewable energy of sensor nodes due to its direct impact on network lifetime dominates every aspect of wireless sensor networks. Existing techniques for optimizing energy consumption are based on exploiting node redundancy, adaptive radio transmission power and topology control. Topology control protocols have a significant impact on network lifetime, available energy and connectivity. In this paper we classify sensor nodes as strong and weak nodes based on their residual energy as well as operational lifetime and propose a Clustering based topology control protocol (CTCP) which extends network lifetime while guarantying the minimum connectivity. Extensive simulations in Java-Simulator (J-Sim) show that our proposed protocol outperforms the existing protocols in terms of various performance metrics life network lifetime, average delay and minimizes energy utilization.

## 1 Introduction

Topology of a Wireless Sensor Network determines the connectivity of the wireless network and profoundly impacts the routing algorithms applied to the network. Topology also influences other important features of the network like resiliency and communication cost between nodes. Current research has established efficient network energy utilization as one of the fundamental research issues in wireless sensor networks. Controlling the topology of the network has emerged as an effective solution to the above problem. Like all other aspects of wireless sensor networks, topology control protocols have to be designed and implemented subject to a severe set of computational and energy constraints.

Topology control protocols are designed to exploit node density in the network to extend the network lifetime and provide connectivity. The following criteria have been identified as the key concepts for designing topology control protocols for wireless sensor networks.

- Sensor nodes should be able to self-configure to accommodate changing network dynamics.

- Selection of redundant nodes should be done based on distributed localized algorithms.
- Topology control protocols must ensure minimum connectivity in the network, so that the network is not partitioned.
- Topology control protocols should take advantage of the high node density in large-scale wireless sensor networks to reduce the energy dissipated in the network.

## 2 Literature Surveys of Topology Control Protocols for Wireless Sensor Networks

Extending network operational lifetime seems to be a key factor in the design of network layer or MAC layer protocols for sensor networks. Topology control protocols can be classified into two groups depending on which network layer information is used for identifying redundant nodes:
- Protocols like CEC [1], GAF [1], ASCENT [2], and LEACH [3] use information from the routing layer and above for identifying redundant nodes.
- Protocols like PAMAS [4] [6], STEM [5] use MAC Layer information to identify redundancy in the network.

GAF and CEC are the two protocols which provide the foundation for our proposed Clustering based topology control protocol (CTCP). Hence GAF and CEC are next discussed and analyzed in greater details.

### 2.1 Geographic Adaptive Fidelity (GAF)

GAF [1] is a location based energy conservation protocol. In GAF redundant nodes are identified based on their geographic locations. The radio of a node is periodically switched off for balancing the load. Location information in GAF is provided by Global Positioning System (GPS) and GAF assumes that the location information is correct. GAF uses the concept of equivalent nodes. Equivalent nodes are intermediate nodes which are same in terms of their connectivity to other nodes with respect to communication. In GAF the network area is divided into small virtual grids such that all nodes in adjacent grids are in each other's radio range. Thus in each virtual grid any one of the nodes can be used for routing. In GAF thus energy saving can be done by keeping the radio of one sensor node active per grid and switching off the radios of all the other sensor nodes. To further balance the energy dissipation in each grid the nodes in a grid are periodically rotated to be active, so that at any given time only one node is switched on per virtual grid.

Issues with GAF:
- GAF is dependent on global information. It fails in applications were geographic location information is not available and hence GAF can be used in very limited applications

- In GAF if a grid has only one node then is not possible to balance the energy usage for that virtual grid and the network may have pockets of low energy virtual grids which in turn may lead to network partitioning.

**2.2 Cluster Based Energy Conservation Protocol (CEC)**

We next studied the Cluster based Energy Conservation (CEC) protocol which directly and dynamically measures the network connectivity so that energy can be conserved by identifying the nodes which can be selectively powered off. CEC [1] configures the nodes into overlapping clusters. CEC also has cluster-head node and gateway node for each cluster for the purpose of maintaining connectivity in the network. CEC tries to provide solutions for some of the problems of GAF. Clustering is one of the most fundamental ways used to design scalable sensor networks. A clustering algorithm arranges the network into subsets of nodes each with a cluster-head at approximately the center of each cluster. A regular high level structure is obtained from good clustering [6]. It is easier to design efficient energy conserving protocols for this kind of structured organization of nodes than at the level of individual nodes. This type of topology also eliminates the need for global information and localized algorithms can then be used in these clusters to reduce the centralized coordination and require that nodes interact with only their neighbors further reducing the communication costs. CEC defines a cluster as a subset of nodes which are mutually reachable in two hops. In CEC cluster formation takes place in a distributed fashion and clusters are interconnected to each other through overlapping nodes. Each cluster has a cluster- head and all the members are within direct radio range of the cluster-head. In CEC the cluster formation takes place in the following manner:
- Initially each node broadcasts discovery messages that contains its node ID, cluster ID and estimated lifetime.
- After receiving discovery messages from all its neighbors if a node sees that it has the highest energy among all its neighbors it declares itself as a cluster-head and broadcasts this.
- If a non-cluster head node receives cluster-head messages from more than one cluster-head it declares itself as a gateway node and broadcasts this information. CEC first selects the cluster-head and then the gateway nodes connecting the clusters. A node elects itself as the cluster-head if it has the longest lifetime among its entire neighbors, breaking ties by node ID. For gateway selection from multiple gateways the gateway with the longest lifetime is assigned highest priority.
- Next leaving the cluster-head and gateway nodes all the other nodes are powered off to conserve energy. After periodic intervals the entire clustering process is reiterated.

We propose CTCP (Clustering based topology control protocol), a clustering protocol which ensures minimum connectivity in the network by optimizing energy consumption and addresses the issue of problems with network connectivity in CEC. In this protocol we propose and define the concept of strong and weak sensor nodes based on their operational lifetime.

# 3 Clustering based topology control protocol (CTCP)

First we define the following:
- A cluster is defined as a set of nodes that are mutually reachable in at most two hops. Each cluster has a cluster-head which is directly reachable from all members of the cluster.
- A gateway is defined as a node which is a member of more than one cluster and provides interconnection between the clusters. In CTCP, during clustering, a node can be in one of the four states, namely, cluster-head, gateway, potential cluster-head, and ordinary node.
- Re-clustering Interval (RI) is defined as the time after which re-clustering is initiated in a cluster.

$$RI = \alpha \; Lifetime \; of \; cluster \; head, where \; 0 < \alpha < 1.$$

Both the value of the estimated lifetime of the cluster-head as well as I change with time. Each cluster has its own re-clustering interval.

- Strong and Weak nodes: A node is defined as a strong node if its lifetime is greater than RI when it operates at full power for the entire duration of its lifetime otherwise it is defined as weak node.

## 3.1 Cluster Formation

Phase 1: cluster-head selection

Initially each node broadcasts a discovery message which contains its node ID, its Cluster ID, and estimated lifetime. A node elects itself as a potential cluster-head if it has the longest lifetime among all its neighboring nodes (ties are broken by node ID). After a node elects itself as a potential cluster head it broadcasts this information along with its lifetime to all its neighbors.

Phase 2: Gateway selection

A node which is directly reachable from more than one cluster-head is called a primary gateway. A node which is connected to the cluster-head of another cluster through a member of that cluster is called a secondary gateway. When a node receives messages from more than one potential cluster-heads it knows that it is a gateway and then decides whether it is a strong node or a weak node with respect to the potential cluster-heads. The gateway node then passes the information whether it is a strong or a weak node to the potential cluster-heads. A gateway node between two potential cluster-heads can thus be a strong node with respect to one potential cluster-head and weak node with respect to the other potential cluster-head. In this case two different messages are sent to the two different potential cluster-heads.

Phase 3: Cluster-head selection and cluster formation

If a potential cluster-head receives the information that all its gateways are strong then it elects itself as the cluster-head and broadcasts this to all its neighbors which set their cluster Id to that of the cluster-head and a cluster is formed. The cluster-head broadcasts the value of re-clustering interval RI to all the members of the cluster. If a potential cluster-head receives the information that one or more of its gateways are weak nodes it elects itself as the cluster- head but while broadcasting this information to its neighbors it reduces the value of $\alpha$ hence reducing the re-clustering interval.

After the cluster formation except for the cluster-head and the gateways all the other cluster members switch off their radios for an amount of time equal to the re-clustering interval to minimize the energy consumption.

Figure 1 shows an example of a cluster formation according to CTCP. Here node 2 and node 3 are strong gateways since their remaining energy will allow them to survive the usual re-clustering interval (when $\alpha = 0.50$) while node 7 is a weak gateway since its remaining energy will not allow it to survive the re-clustering interval RI if $\alpha = 0.50$.

In CTCP, an alternative approach can be once a gateway discovers that it is a weak node with respect to a cluster-head it sends intermediate node search messages containing its cluster ID to all its neighbors at some fraction of its maximum transmission power (minimum transmission power is preferred so that the node can last for a longer time [7] [9]). If it gets a reply from its neighbors having the same cluster ID it then uses this neighbor as a bridge node between the cluster-head and itself so that it can be operational for a longer duration of time and extend re-clustering interval. Clustering is expensive in terms of network resources and the number of re-clustering should be kept to a minimum.

### 3.2 Analysis of CTCP

We next analyze CTCP in terms of message complexity as communication between nodes consumes significant energy. For the problem of cluster formation in CTCP the following network model is assumed.

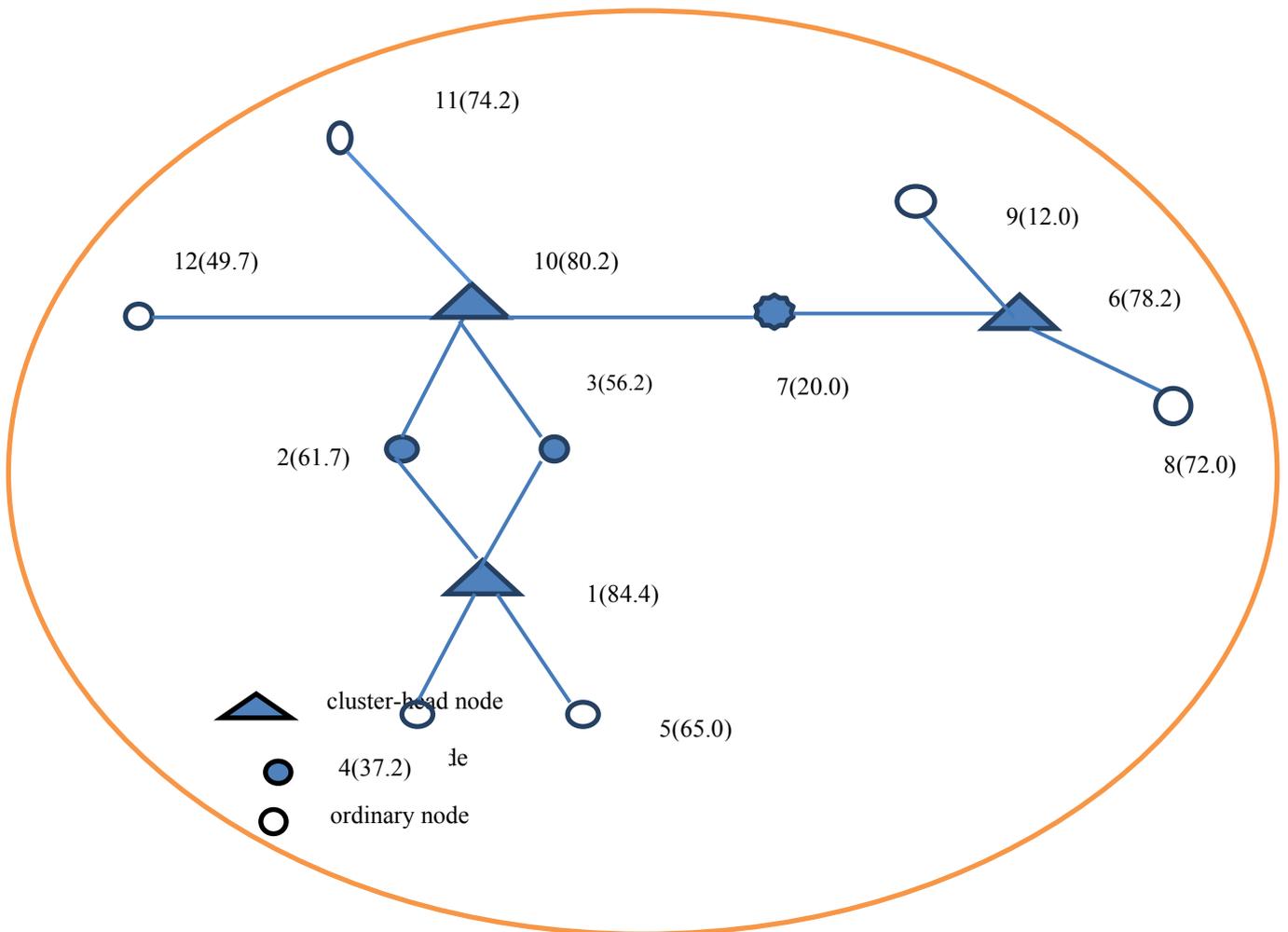

**Fig. 1.** Example of clustering

- A node can directly communicate with its cluster-head.
- A cluster-head has at least one neighbor which is a gateway.
- An ordinary node has at least one neighboring node (a node within direct communication range) which is a cluster-head.

### 3.2.1 Conditions that should be satisfied by the clustering protocol

Based on the above assumptions, our cluster formation algorithm of protocol should satisfy the following criteria similar to the desirable clustering condition in [8] [10].

- Cluster formation is completely distributed. Each node independently chooses to join a cluster based on local information.
- The cluster formation should terminate within a fixed number of iterations.

- By the end of one cluster formation period (FPc) a node is either a cluster-head, a gateway or an ordinary node. A gateway node can at the most belong to the number of clusters given by the number of its neighbors denoted by (k). An ordinary node will belong to only on cluster.

- Clustering should be efficient in terms of number of messages exchanged.

- Inter-cluster connectivity is assured.

### 3.2.2 Correctness Proof and Complexity

In this section we prove the correctness of CTCP and also show that it satisfies all the criterion of a good clustering algorithm. We assume that x denotes the total number of nodes in the network and the y denotes the number of neighbors (within direct communication range) of each node. Note that CTCP is a completely distributed algorithm. A node in CTCP can either become a cluster-head or join a cluster according to received messages. The decision whether to become a cluster-head or a gateway or an ordinary node is based on local information.

**Lemma 1:** By the end of the phase 3 of CTCP, a node is either a cluster-head, a gateway or an ordinary node.

**Proof:** Let us suppose that CTCP terminates and a node is neither of a cluster-head, a gateway or an ordinary node. Then by the definition of states of a node in CTCP, the node must be a potential cluster-head. By the end of Phase 1, of CTCP a node may become a potential cluster-head. By the end of Phase 2, a node may still be a potential cluster-head. But by the end of Phase 3, of CTCP the cluster-head selection in the network is completed. Hence a node which is not a cluster-head or a gateway or an ordinary node by Phase 3, by Assumption 3 of 4, will have a neighbor which is a clustered and hence its status will change from potential cluster-head to an ordinary node. This contradicts our assumption proving the lemma.

**Lemma 2:** One clustering interval of CTCP finishes in O(1) iterations.

**Proof:** From Lemma 1 we see that at the end of one iteration cluster formation is complete and each node is either a clustered or a gateway or an ordinary node. Hence one clustering interval CTCP terminates in constant number of iterations.

**Lemma 3:** In CTCP the message complexity of one clustering interval per node is O(y), and the message complexity of one clustering interval for the whole network is O(x2).

**Proof:** By the end of one complete round of clustering, let $xp_{ch}$ denote the number of potential cluster-heads, $x_{gw}$ denote the number of gateways and $x_{ch}$ denote the number of cluster-heads in the network.

- The total number of messages exchanged in Phase 1 in the network due to the discovery and potential cluster-head messages is $(x * y) + (xp_{ch} * y)$. In the worst case when the graph is fully connected, $x = y$ and the complexity becomes $O(x^2)$.

- Total number of messages exchanged in Phase 2 due to gateway messages is $x_{gw} * y$

- Total number of messages exchanged in the network in Phase 3 for cluster-head notification is $x_{ch} * y$.

Thus for a single clustering interval of CTCP the message complexity is $O(x^2)$.

**Lemma 4:** CTCP ensures minimum connectivity in the network once the clusters are formed.

**Proof:** By Phase 3, of CTCP the clusters are formed. Further, phase 3 of CTCP algorithm ensures that gateways are alive for the entire period of time till the next clustering takes place. Hence minimum network connectivity is ensured in CTCP.

## 4 Simulations of CTCP

The primary objective of CTCP is to optimize the energy conservation in the network while ensuring guaranteed connectivity of the network. We initially simulated our protocol with a small network consisting of a few nodes. The results indicated that CTCP performed better than CEC in providing network connectivity. We next implemented our protocol (CTCP) in (J-Sim) Java Simulator. We ran GAF, CEC and CTCP for the same simulation scenarios to compare their performance. The nodes move randomly with a speed within 0 to 20m/s in a 1500m by 300m area. For simulating a wireless sensor network 50 nodes are used to route data while 10 nodes are used as sources and sinks. The traffic generated was characterized by constant bit rate (CBR). The packet size was set to 512 bytes. The 10 nodes generating the traffic data were given infinite amount of energy. These nodes just generate data and do not participate in data routing. The fifty nodes used for routing were assigned an initial energy values to keep them alive for 500s.

We next evaluate the performance of CTCP with respect to performance metrics like network connectivity, energy consumption and network operational lifetime etc.

**4.1 Network Operational Lifetime**

We first study the survivability of the nodes in all of the three protocols CTCP, GAF and CEC. From figure 2 we see that for a simulation time of 1000s in CTCP on an average 36−37% of more nodes are alive when compared to GAF. In CEC the number of alive nodes is equivalent to CTCP. The re-clustering interval in CTCP is reduced for clusters with weak gateways. Clustering is expensive in terms of network resources due to the messages exchanged for cluster formation.

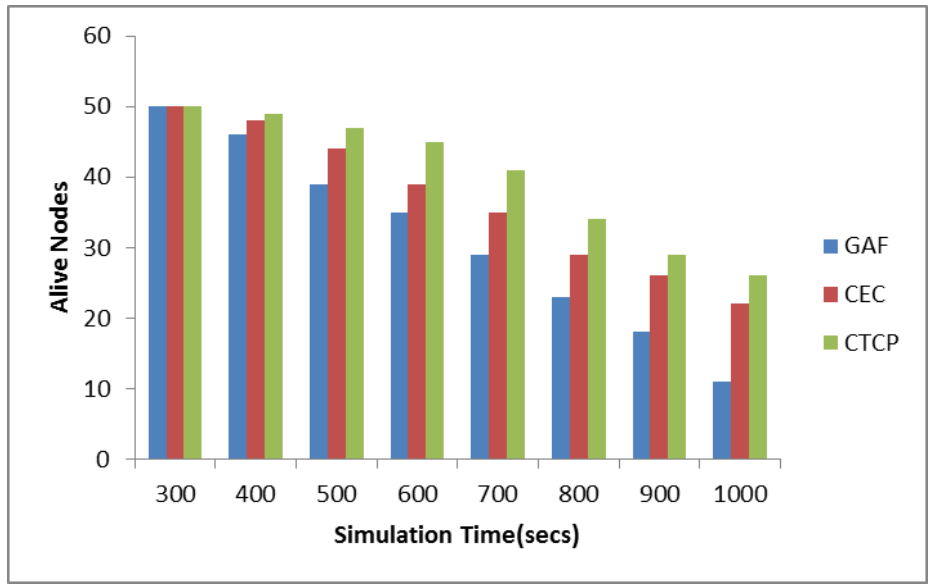

**Fig. 2.** The number of alive nodes with the simulation time

### 4.2 Residual Node Energy

To obtain the average energy of the network at a given simulation time we consider the summation of the energy of all the nodes at that instant of time. From figure 3 we see that the average network energy in CEC and CTCP are much higher than in GAF. CEC shows 20% more average energy than CTCP. The nodes in the network display varying residual energy levels after being operational for some period of time. In case of CTCP to ensure connectivity, re-clustering takes place much more frequently. The increased average network energy in case of CEC comes at the cost of partitions in the network.

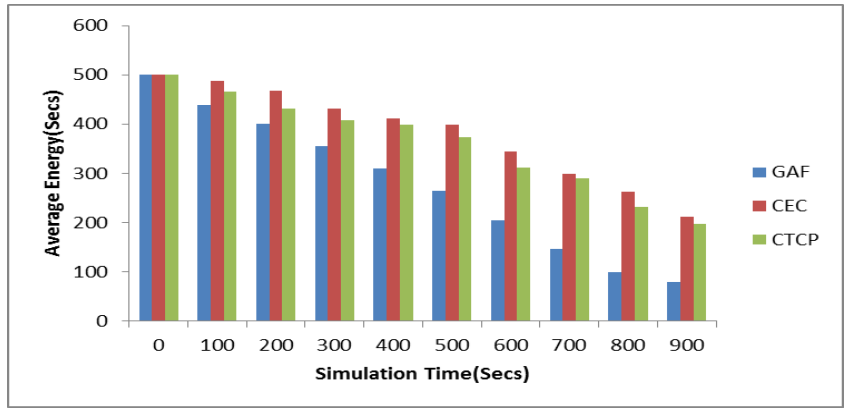

**Fig. 3.** Average energy of the network with simulation time

### 4.3 Average Delay

We use average data delivery delay defined as the mean delay of received packets to measure the connectivity in the network. From figure 4 we see that GAF shows better data delivery ratio at higher mobility. At high mobility the nodes are constantly moving in and out of the virtual grids and thus the energy dissipation of the nodes is balanced, but at low mobility as the node density decreases with time the connectivity in the network is accepted in case of GAF. Here we use pause time to indicate mobility of the nodes. Pause time of 0s indicates that the nodes are constantly moving while a pause time of 1000s indicates that the nodes are almost static. The spikes in the graph of CEC denote network partitions which are caused when gateway nodes run out of energy. No such spikes are observed for CTCP as CTCP tracks the gateway node residual energy and re-clusters before the connectivity is broken. In CTCP the density of nodes is exploited to conserve energy. Thus with higher node density, even more energy savings can be achieved.

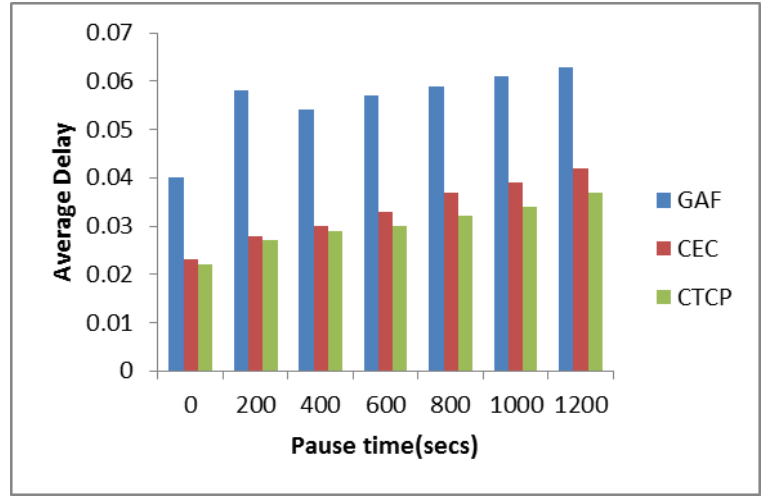

**Fig. 4**. Average delay with pause time

**4.4 Trade-off between Network Connectivity and Residual Energy of the Nodes**

From all of the above figures we see that the performance of CTCP and CEC are comparable in terms of performance metrics like network operation lifetime, average energy of the network etc., but CTCP outperforms CEC with respect to connectivity. Hence CTCP is more suitable for critical real-time applications than CEC. Comparing CTCP and GAF we see that CTCP outperforms GAF in terms of all the performance metrics. The above simulation clearly establishes the existence of trade-off between network connectivity and energy conservation in the network based on a given topology. The optimal value of this trade-off can be set by the specific application in question.

## 5 Conclusion

Topology control protocols determine the energy consumption as well as connectivity of the network. CTCP is a distributed localized protocol independent of the routing protocol used in the network. CTCP exploits the node redundancy in large scale dense sensor networks for network energy conservation. CTCP classifies nodes as strong and weak based on their residual energy and uses this classification to achieve a balanced distribution of energy across the network. CTCP organizes the topology of the network into overlapping clusters, elects a cluster-head and gateway nodes for each cluster. Next all other nodes in the cluster except the cluster-head and gateway are identified as redundant nodes and powered off. The underlying principle in CTCP is to minimize the power consumption of the radio in sensor nodes. Previous research has proved that radio is the primary source of energy dissipation in sensor networks. CTCP provides ensured connectivity in the network via gateway nodes. In our proposed protocol the decision of re-clustering is based on the residual energy of both the cluster-head and gateway nodes. In CTCP further energy optimizations are achieved by powering of redundant gateways. As the process of cluster formation is message intensive CTCP invokes clustering only after the nodes show unequal distribution of residual energy. We simulated CTCP in Java Simulator (J-Sim) and evaluated the performance of CTCP in terms of various performance metrics such as average network energy, network operational lifetime, connectivity. CTCP clearly displays superior connectivity and network energy optimization with respect to existing protocols like CEC and GAF.